\documentclass[a4paper]{jpconf}
\usepackage{graphicx}
\usepackage{amsmath}
\begin{document}
\title{Regular and chaotic motion in general relativity: The case of an inclined black hole magnetosphere}

\author{Ond\v{r}ej Kop\'{a}\v{c}ek and Vladim\'{i}r Karas}

\address{Astronomical Institute, Academy of Sciences, Bo\v{c}n\'{i}~II~1401/1a, CZ-141\,31~Prague, Czech~Republic}

\ead{kopacek@ig.cas.cz}

\begin{abstract}

Dynamics of charged particles in the vicinity of a rotating black hole embedded in the external large-scale magnetic field is numerically investigated. In particular, we consider a non-axisymmetric model in which the asymptotically uniform magnetic field is inclined with respect to the axis of rotation. We study the effect of inclination onto the prevailing dynamic regime of particle motion, i.e. we ask whether the inclined field allows regular trajectories or if instead, the deterministic chaos dominates the motion. In this contribution we further discuss the role of initial condition, particularly, the initial azimuthal angle. To characterize the measure of chaoticness we compute maximal Lyapunov exponents and employ the method of Recurrence Quantification Analysis.
\end{abstract}

\section{Introduction}
In this paper we investigate some aspects of dynamics of ionised matter exposed to the interplay of strong gravitational and electromagnetic fields. In particular, we consider the astrophysically motivated model consisting of a rotating Kerr black hole that is immersed in the asymptotically uniform magnetic field. Previously we have investigated dynamics of charged particles in the axisymmetric case with parallel orientation of the field with respect to the rotation axis \cite{kovar10, kopacek10, kopacek10b}. We studied the topology of the particle confinements and located both the regions of mostly regular motion as well as areas dominated by the deterministic chaos. 

More recently we have also investigated the generalized scenario in which the magnetic field is arbitrarily inclined with respect to the axis \cite{kopacek14}. Breaking the axial symmetry appears to have profound consequences regarding the dynamics of matter. Most importantly, we found that within the given model the stability of regular motion of charged particles depends critically on the perfect alignment of large-scale magnetic field with the rotation axis. Once the field is slightly inclined and the symmetry of the system breaks, the chaotic regime completely dominates the dynamics. We quantified the chaoticness of orbits by the maximal Lyapunov exponent $\chi$. Regarding the trajectories which are regular in the axisymmetric setup we found that inclining the field causes almost immediate onset of chaos. Increasing the inclination angle typically leads to the increase of $\chi$ until it saturates. On the other hand, if we consider trajectories which already show distinctive chaotic features in the aligned field, we find that $\chi$ is almost indifferent to the increasing inclination. Obliqueness of the field, which may be considered as an extra perturbation of the system, does not enhance chaos in this case.

Naturally, breaking the axial symmetry cancels the conservation of the axial component of the angular momentum $L_z$ and azimuthal angle $\varphi$ is no longer a cyclic coordinate. Therefore, it becomes dynamically important and, especially, by setting different initial azimuthal angles $\varphi(0)$ we obtain fundamentally distinct trajectories. Nevertheless, we shall only consider small inclinations of the field (typically up to $\approx6^{\circ}$). In this context we investigate to which degree the dynamics of test particles may depend on the choice of initial azimuthal coordinate $\varphi(0)$. In particular, we shall explore if different value of $\varphi(0)$ may substantially change the degree of chaoticness.

\section{Rotating black hole in the external magnetic field}
Kerr metric  describing the geometry of the spacetime around the rotating black hole of mass $M$ and spin $a$ may be expressed in Boyer-Lindquist coordinates $x^{\mu}= (t,\:r, \:\theta,\:\varphi)$ as follows \cite{mtw}:
\begin{equation}
\label{metric}
ds^2=-\frac{\Delta}{\Sigma}\:[dt-a\sin{\theta}\,d\varphi]^2+\frac{\sin^2{\theta}}{\Sigma}\:[(r^2+a^2)d\varphi-a\,dt]^2+\frac{\Sigma}{\Delta}\;dr^2+\Sigma d\theta^2,
\end{equation}
where
\begin{equation}
{\Delta}\equiv{}r^2-2Mr+a^2,\;\;\;
\Sigma\equiv{}r^2+a^2\cos^2\theta.
\end{equation}
Geometrized units are used throughout the paper. Values of basic constants thus equal unity $G=c=k=k_C=1$.

A test-field solution corresponding to the aligned magnetic field (of the asymptotic strength $B_z$) was derived by Wald \cite{wald74}. This solution was later generalized by Bi\v{c}\'{a}k and Jani\v{s} \cite{bicak85} to describe the field which is arbitrarily  inclined with respect to the rotation axis (specified by two independent components $B_z$ and $B_x$). We may also consider a non-zero electric charge $Q$ of the black hole generating the electromagnetic field of Kerr-Newman black hole, though in the test-field regime (metric remains unaltered by $Q$). Resulting vector potential of the electromagnetic field $A_{\mu}=(A_t,A_r,A_{\theta},A_{\varphi})$ can be given explicitly \cite{kopacek14}.   

We note that in our previous studies \cite{karas12,karas09} we have investigated the structure of the electromagnetic field arising in this setup enriched by an extra ingredient: uniform motion of the black hole in arbitrary direction with respect to (arbitrarily inclined) magnetic field. However, here we consider the inclination of the magnetic field but not the boost of the black hole.

\subsection{Equations of motion}
Employing the Hamiltonian formalism we first construct the super-Hamiltonian $\mathcal{H}$ \cite{mtw}:
\begin{equation}
\label{hamiltonian}
\mathcal{H}=\textstyle{\frac{1}{2}}g^{\mu\nu}(\pi_{\mu}-qA_{\mu})(\pi_{\nu}-qA_{\nu}),
\end{equation}
where $q$ is charge of the test particle (of rest mass $m$), $\pi_{\mu}$ is the generalized (canonical) momentum, $g^{\mu\nu}$ is the metric tensor, and $A_{\mu}$ denotes the vector potential. The latter is related to the electromagnetic tensor $F_{\mu\nu}$ by $F_{\mu\nu}=A_{\nu,\mu}-A_{\mu,\nu}$. 

Hamilton's equations of motion are given as
\begin{equation}
\label{hameq}
\frac{{\rm d}x^{\mu}}{{\rm d}\lambda}\equiv p^{\mu}=
\frac{\partial \mathcal{H}}{\partial \pi_{\mu}},
\quad 
\frac{d\pi_{\mu}}{d\lambda}=-\frac{\partial\mathcal{H}}{\partial x^{\mu}},
\end{equation}
where $\lambda=\tau/m$ is the affine parameter (dimensionless in geometrized units), $\tau$ denotes the
proper time, and $p^{\mu}$ is the standard kinematical four-momentum for
which the first equation reads $p^{\mu}=\pi^{\mu}-qA^{\mu}$ and thus the conserved value of super-Hamiltonian is equal to $\mathcal{H}=-m^2/2$.
Moreover, the system is stationary since the Hamiltonian is independent on the coordinate time $t$. Its conjugate momentum $\pi_t$ is therefore an integral of motion. Namely, it expresses (negatively taken) energy of the test particle $\pi_t\equiv-E$. The mass of the black hole $M$ is used to scale all quantities which is formally equivalent to setting $M=1$ in the equations. We also switch to specific quantities $q/m\rightarrow q$ and $E/m\rightarrow E$ when describing the test particle.

\begin{figure}[htb]
\centering
\includegraphics[width=91mm]{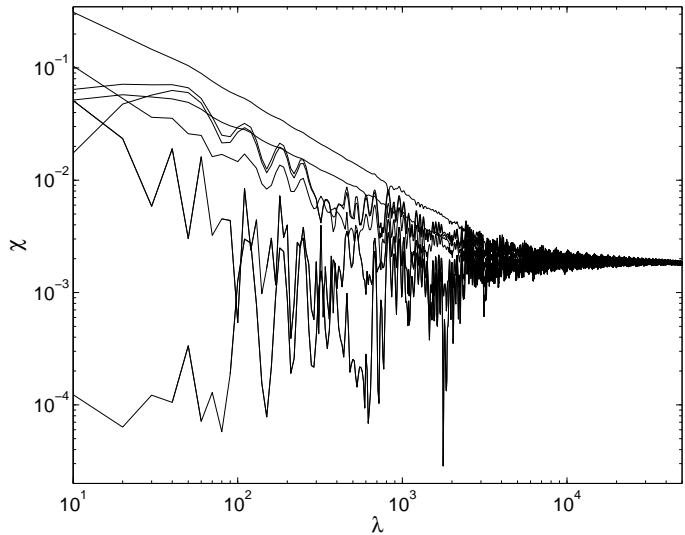}\hspace{5mm}
\begin{minipage}[b]{60mm}\caption{\label{lyanez}Independence of the asymptotical value of the Lyapunov exponent with respect to the choice of the initial deviation $w(0)$ for a particular choice of the chaotic trajectory. The initial deviation with a single non-zero component (e.g., $w(0)=(0,0,0,1,0,0,0,0)$ pointing in the azimuthal direction) results in wildly oscillating $\chi$ while for the `averaged' initial deviation $w(0)=1/\sqrt{8}\left(1,1,1,1,1,1,1,1\right)$ the evolution is rather smooth (uppermost curve in this figure). 
}
\end{minipage}
\end{figure}

\subsection{Maximal Lyapunov characteristic exponent}
Maximal Lyapunov exponent $\chi$ is commonly used as a basic quantitative indicator of chaotic dynamics \cite{lieberman92}. Its value directly captures the tendency of nearby orbits to diverge as the system evolves. In other words, it allows us to express how sensitive the given orbit is on the initial condition. High (exponential) sensitivity is a defining property of chaos. Exponent $\chi$ is defined as follows \cite{skokos10} \footnote{Lyapunov exponent $\chi$ is defined as an asymptotic measure (\ref{mle}). In the numerical application we actually  compute the quantity usually denoted as {\em finite time Lyapunov exponent}, which depends on the integration variable, instead of the limit. However, in this text we disregard such distinction and use the term Lyapunov exponent for both quantities as it cannot cause any confusion within the scope of the paper.}

\begin{equation}
 \label{mle}
\chi\equiv\lim_{\lambda\to\infty}\frac{1}{\lambda}\ln\frac{||w(\lambda)||}{||w(0)||},
\end{equation}
where we choose to use standard $L^2$ (Euclidean) norm to measure the length of deviation vector in the phase space $w(\lambda)=(\delta t,\delta r, \delta \theta, \delta \varphi, \delta\pi_t, \delta\pi_r, \delta\pi_\theta, \delta\pi_\varphi)$. 

\begin{figure}[ht!]
\centering
\includegraphics[scale=0.47, clip]{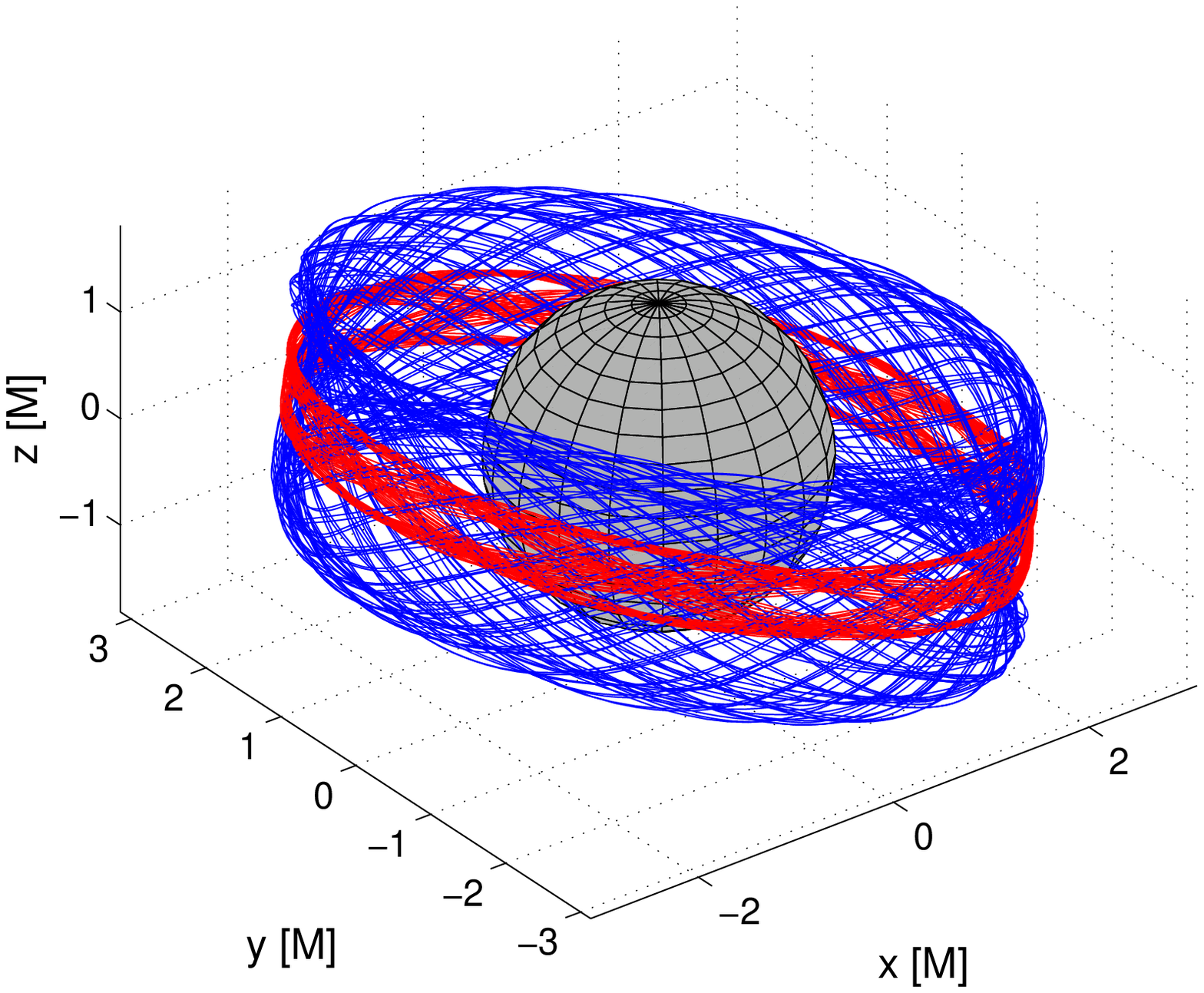}~
\includegraphics[scale=0.44, clip]{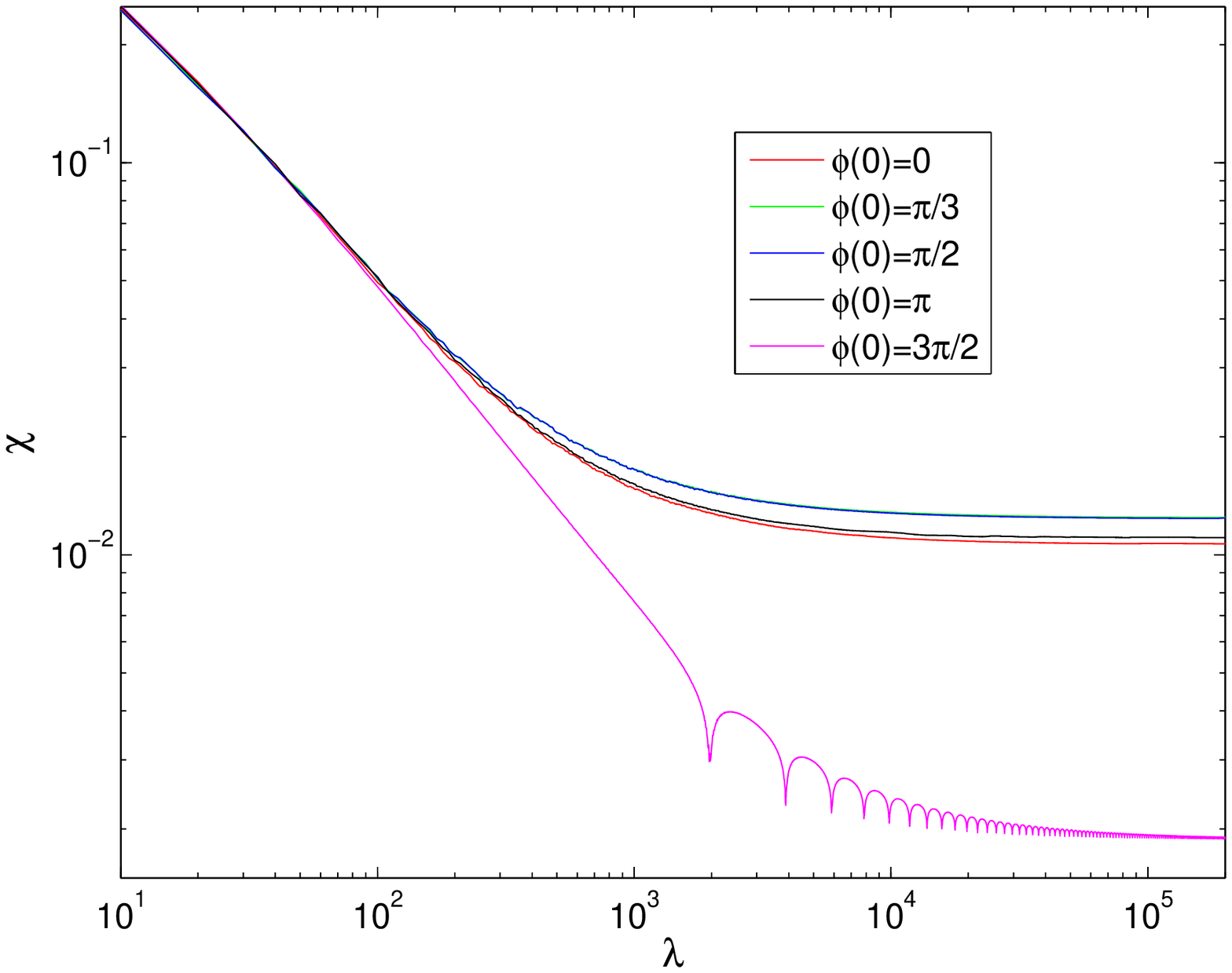}
\includegraphics[scale=0.76, clip]{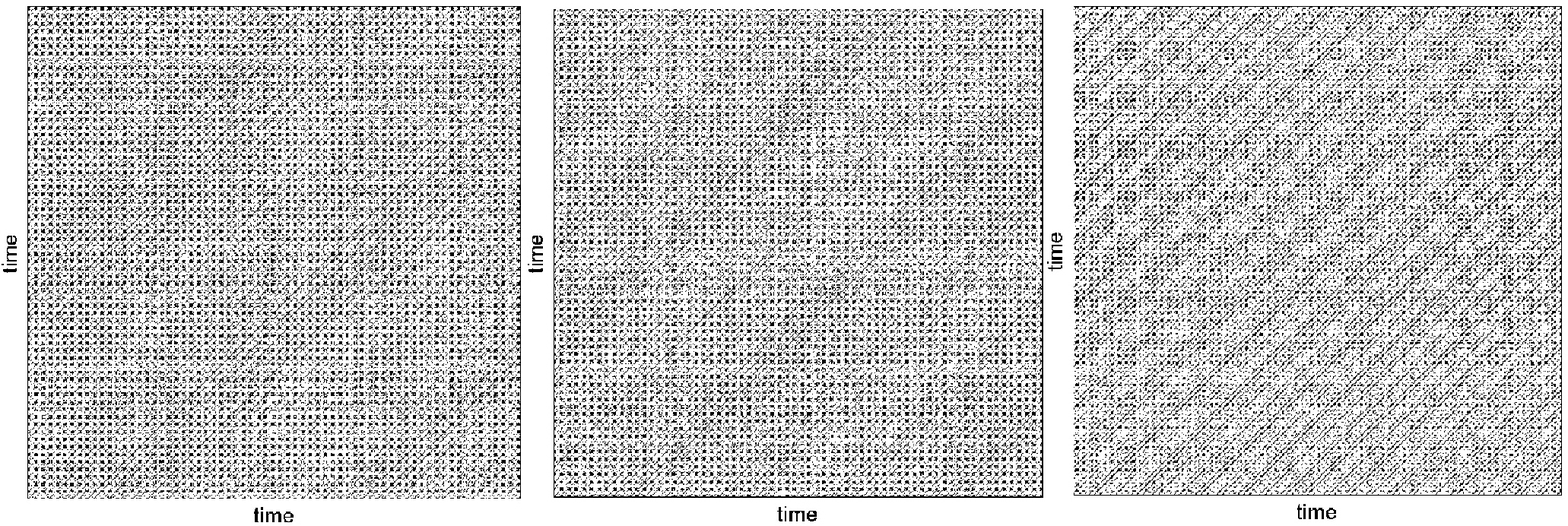}
\caption{The top left panel shows the three-dimensional view of two chaotic orbits in the oblique magnetic field differing solely in the initial azimuthal angle; $\varphi(0)=0$ (blue trajectory) and $\varphi(0)=3\pi/2$ (red orbit). Grey surface marks the horizon of the black hole. In the top right panel we compare Lyapunov exponents of these trajectories which only differ in $\varphi(0)$. Bottom panels compare these trajectories by means of the recurrence plots. The left plot ($\varphi(0)=0$) shows no differences from the middle one ($\varphi(0)=\pi/3$) while the case $\varphi(0)=3\pi/2$ (bottom right) produces slightly different pattern.}
\label{diskuze_phi1}
\end{figure}

We employ the usual method of determining the evolution of the deviation $w(\lambda)$ consisting in solving variational equations, which restrict to the linear term in corresponding Taylor expansion \cite{kaltchev13}. We set the initial deviation as follows $w(0)=1/\sqrt{8}\left(1,1,1,1,1,1,1,1\right)$. Theory of Lyapunov spectra guarantees that setting random initial deviation results in the maximal exponent $\chi$ with probability one. The set of initial deviations for which we would obtain different Lyapunov exponent has zero measure \cite{skokos10}. In Fig.~\ref{lyanez} we illustrate the independence of the asymptotic value of $\chi$ on a particular choice of the chaotic orbit.

\section{Dynamics of charged particles in the oblique field}
We address the question of azimuthal dependence of charged particle dynamics in the inclined magnetic field. The system of test particle in this non-axisymmetric setup has three degrees of freedom and is apparently non-integrable. In general, we expect to find both regular and chaotic orbits in such a system. Our previous analysis \cite{kopacek14} shows that most regular orbits found in the axisymmetric case are quickly destroyed once the field inclines even very slightly. We found that a very small inclination angle is sufficient for the deterministic chaos to dominate the phase space completely. We demonstrate this fact on a series of examples and treat separately several distinct classes of trajectories. Nevertheless, in \cite{kopacek14} we do not discuss the role of initial azimuthal coordinate $\varphi(0)$ which becomes relevant if the axisymmetry is lost.

\begin{figure}[!ht]
\centering
\includegraphics[scale=0.29, clip]{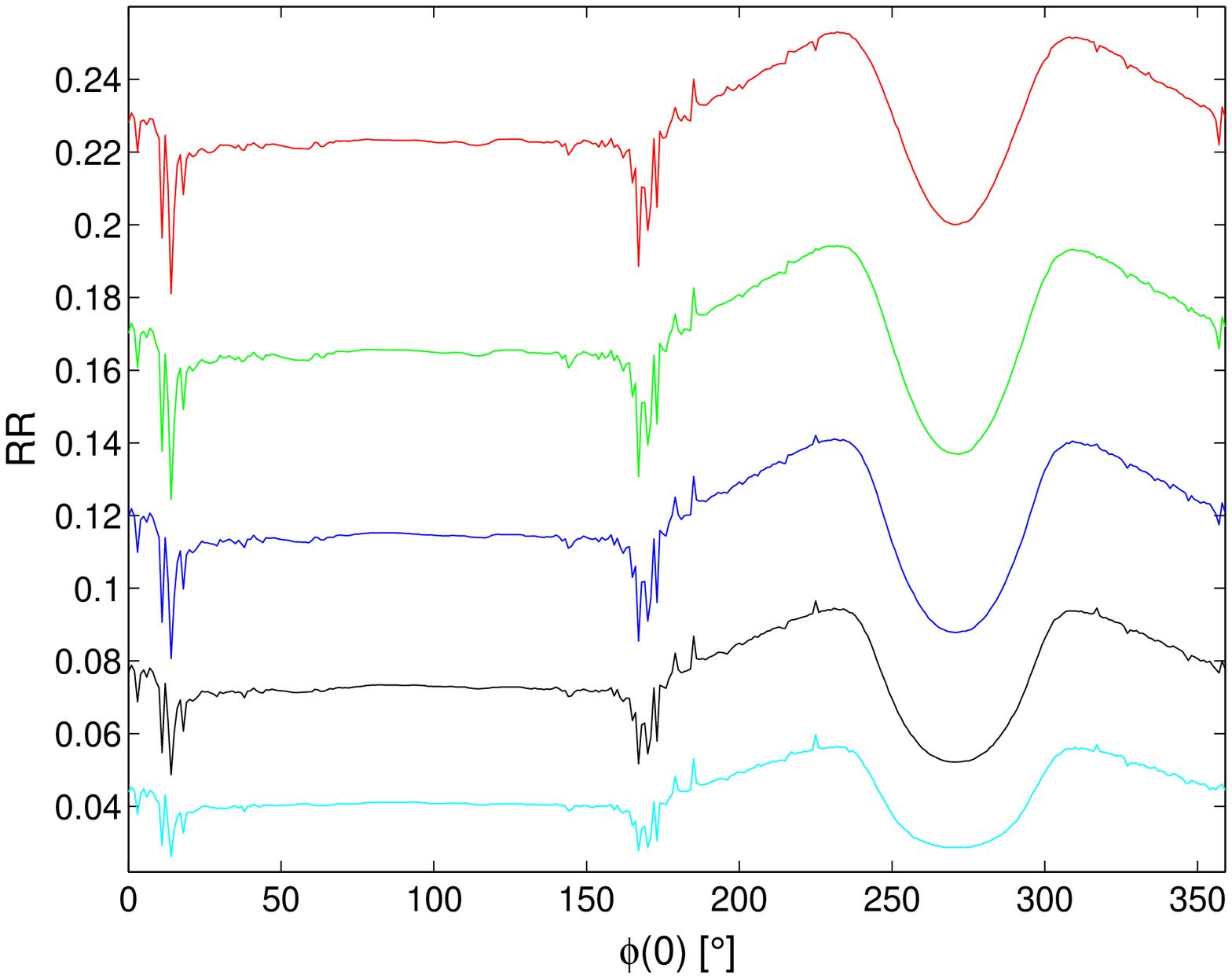}
\includegraphics[scale=0.29, clip]{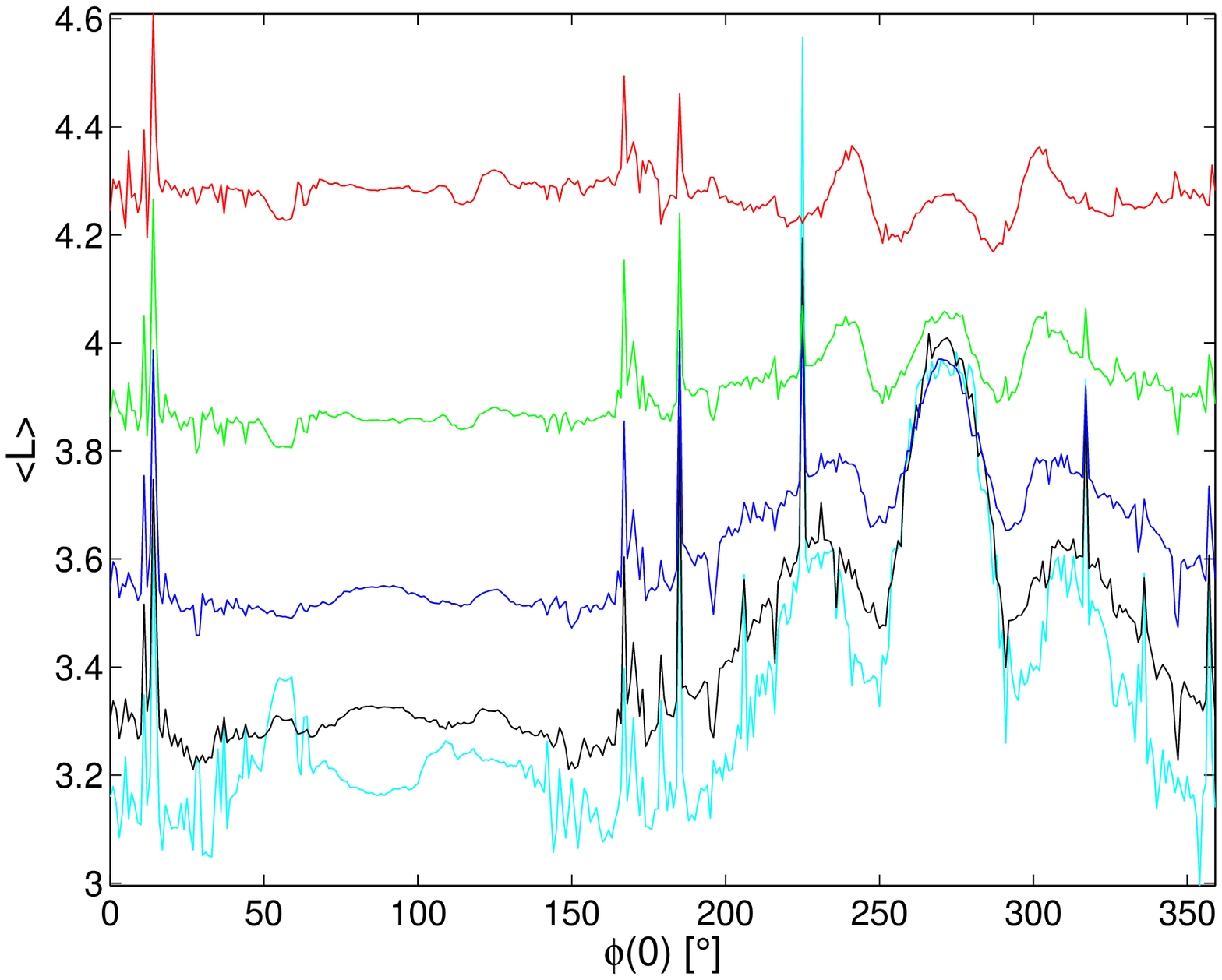}
\includegraphics[scale=0.29, clip]{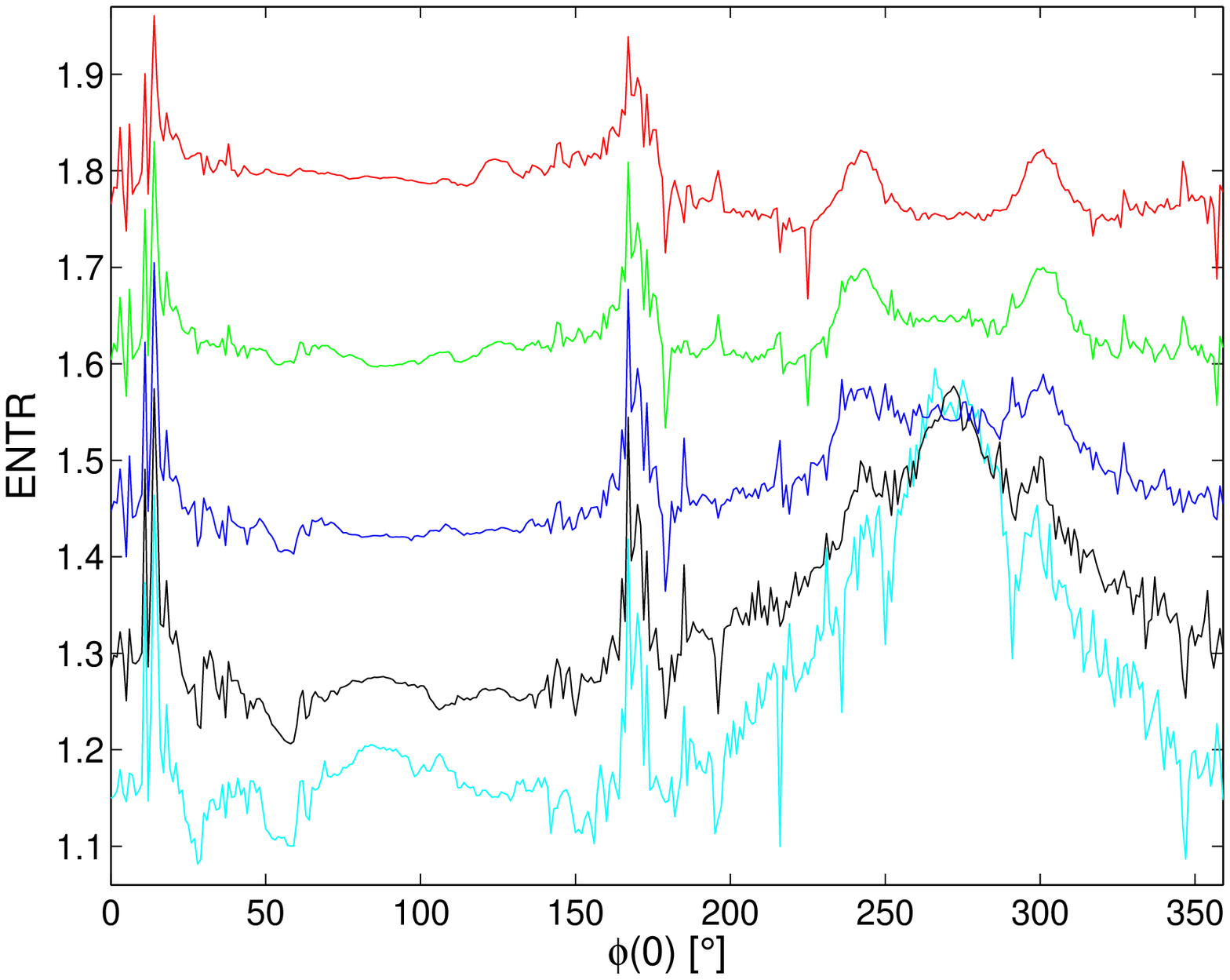}
\includegraphics[scale=0.29, clip]{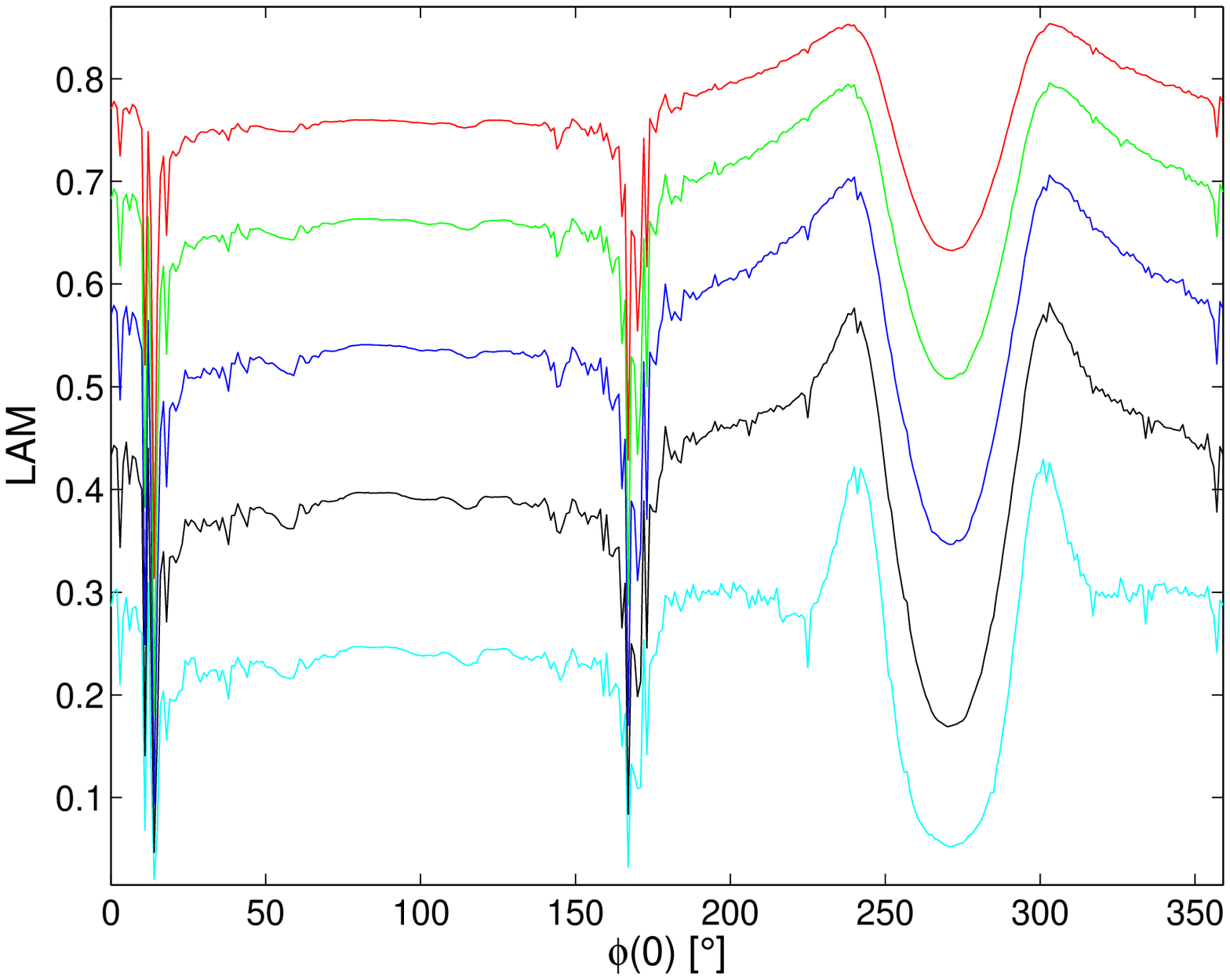}
\includegraphics[scale=0.29, clip]{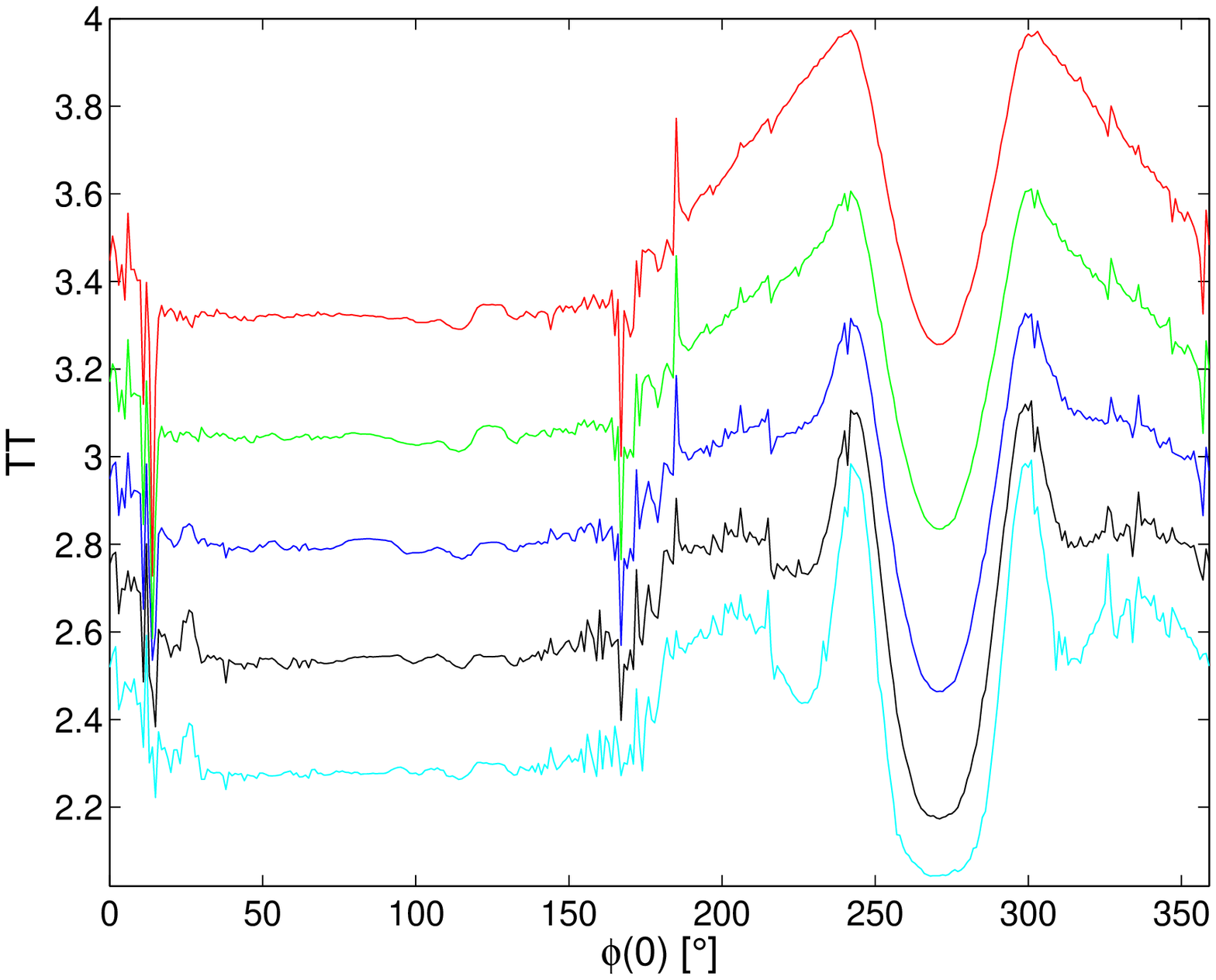}
\includegraphics[scale=0.29, clip]{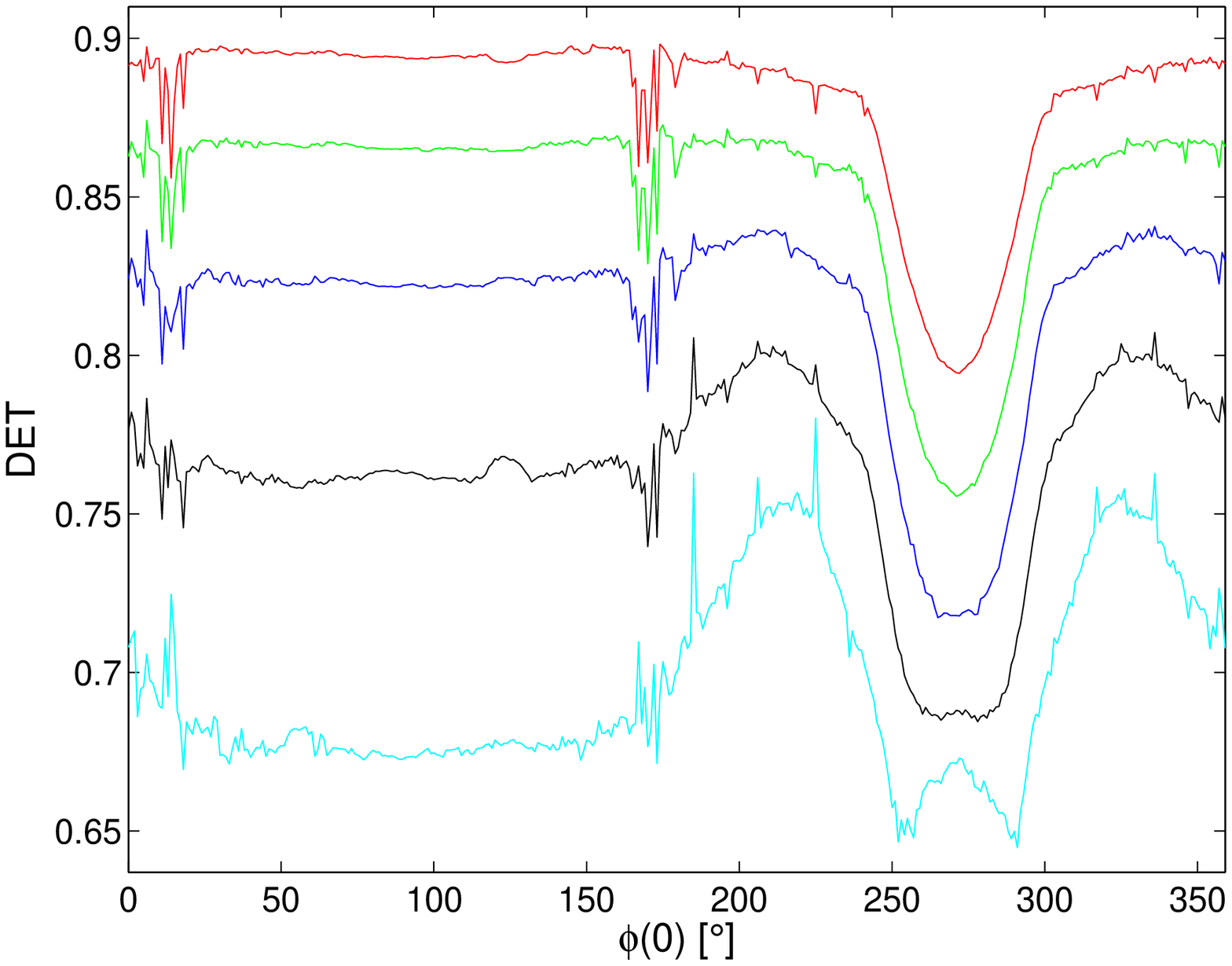}
\caption{Several RQA measures as a function of $\varphi(0)$ (with step $\Delta\varphi=1^{\circ}$) of trajectories analyzed in Fig.~\ref{diskuze_phi1} are shown. Each of them is computed for five distinct values of the recurrence threshold $\varepsilon$ (red: $\varepsilon=2.2$, green: $\varepsilon=1.95$, blue: $\varepsilon=1.7$, black: $\varepsilon=1.45$, cyan: $\varepsilon=1.2$) with  time series of each coordinate normalized separately to zero mean and standard mean deviation equal to unity (see \cite{marwan07} for details). }
\label{diskuze_phi_rqa}
\end{figure}

We shall not present any systematic discussion of the topic here. We rather demonstrate the effect of different initial azimuthal angle on a suitably chosen example. To this end, we pick the chaotic trajectory specified by following values of parameters $a=0.9$, $E=1.58$, $qQ=0$, $qB_z=1$, $B_x/B_z=0.1$ and with the initial condition set as follows: $r(0)=3$, $\theta(0)=\pi/2$, $u^r(0)=0$ and $\pi_{\varphi}(0)=5$. This particular set of values was used in our previous analysis (see Fig.~4 in \cite{kopacek14}) in the discussion of the impact of gradually increasing inclination on the trajectory which is regular in the parallel field ($B_x/B_z=0$). However, here we choose the highest value of inclination permitted for given parameters ($B_x/B_z=0.1$ corresponding to angle $\approx6^{\circ}$) in order to make the possible effect of different $\varphi(0)$ most apparent. 

In Fig.~\ref{diskuze_phi1} we show that most values of $\varphi(0)$ lead to similar results in terms of Lyapunov exponent $\chi$. Indeed, most angles result in a similar evolution as well as the final value of $\chi$, while only the case of $\varphi(0)=3\pi/2$ leads to a qualitatively different behaviour and $\chi$ converges to the significantly smaller value. With $\varphi(0)=3\pi/2$ the trajectory visibly differs from others also in direct comparison of their three-dimensional structures (top left panel of Fig.~\ref{diskuze_phi1}). To further explore and compare its dynamics we employ the technique of recurrence quantification analysis (RQA) \cite{marwan07}. First we construct recurrence plots (RPs) of trajectories with different values of $\varphi(0)$. In bottom panels of Fig.~\ref{diskuze_phi1} we compare RPs for $\varphi(0)=0$, $\varphi(0)=\pi/3$ and $\varphi(0)=3\pi/2$. The former two RPs are almost identical while the latter has slightly different and more diagonally oriented pattern which is a general hallmark of less divergent dynamics. Thus the visual survey of RPs is in agreement with the observation that $\varphi(0)=3\pi/2$ leads to significantly lower value of $\chi$, i.e. is less chaotic in this sense. However, recurrences of the trajectory in the phase space may also be quantified in terms of various RQA measures (see \cite{marwan07} for their definitions). In Fig.~\ref{diskuze_phi_rqa} we present plots of several of these indicators as a function of $\varphi(0)$. We observe that profound change of dynamics for $\varphi(0)=3\pi/2$ is detected by all these recurrence measures. Although the interpretation of RQA measures is not straightforward, we note that RQA represents a suitable tool to systematically search the parameter space for any anomalies in resulting dynamics. In particular, here it proves that the case $\varphi(0)=3\pi/2$ is dynamically distinct from all other values of $\varphi(0)$. 

\begin{figure}[htb]
\centering
\includegraphics[width=91mm]{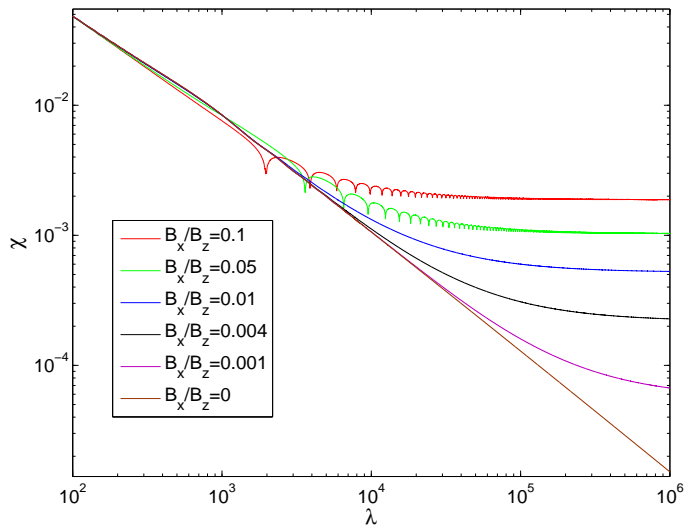}\hspace{5mm}
\begin{minipage}[b]{60mm}\caption{\label{diskuze_Bx}Comparison of Lyapunov exponents of trajectories with $\varphi(0)=3\pi/2$ which only differ in inclination of the field. For the aligned field the trajectory remains regular, however, if the field even only slightly inclines the chaos sets on immediately. Then the gradual growth of the largest Lyapunov exponent $\chi$ and also the emergence of damped oscillations is observed as the inclination of the field increases. Logarithmic scale is used for the both axes.}
\end{minipage}
\end{figure}

Finally, in Fig.~\ref{diskuze_Bx} we further investigate the case $\varphi(0)=3\pi/2$ and calculate the evolution of $\chi$ for several values of inclination. We observe that with decreasing $B_x/B_z$ the extraordinary behaviour represented by damped oscillations of $\chi$ quickly disappears and the dynamics gradually approaches regular behaviour which is reached for zero inclination. Ordered sequence of asymptotical values of $\chi$ with respect to the inclination angle confirms the results of our previous analysis \cite{kopacek14} which was perfomed for the trajectory launched at $\varphi(0)=0$.

\section{Conclusions}
We have numerically analyzed the role of initial azimuthal angle in determining the dynamic properties of charged particles in slightly (but not negligibly) non-axisymmetric setup of rotating black hole immersed in the oblique magnetic field. Demonstrating its effect on the suitably chosen example we have seen that even for small inclination of the field, the choice of $\varphi(0)$ may lead to substantially different behaviours of particles. Employing the largest Lyapunov exponent $\chi$ and the recurrence quantification analysis, we have shown that although most cases may lead to the similar results, we can also find some prominent mutual orientation of the field and the initial velocity of the particle which produces markedly different trajectory with distinct evolution and significantly different asymptotic value of $\chi$. This change of dynamics was also clearly detected by various RQA measures. We conclude that even in a weakly non-axisymmetric system the role of initial azimuthal angle becomes highly important and its effect cannot be neglected.

\ack
OK acknowledges the postdoctoral program of the Czech Academy of Sciences. VK thanks the Czech Science Foundation for support via the project GA\v{C}R 14-37086G.

\section*{References}
\bibliography{ere2014}

\end{document}